\def \be {\begin{equation}}
\def \eq {\end{equation}}
\def \bee {\begin{eqnarray}}
\def \eqq {\end{eqnarray}}

\def \bea {\begin{array}{c}}
\def \eqa {\end{array}}

\def \dels {\partial\kern-.5em / \kern.5em}
\def \As {{A\kern-.5em / \kern.5em}}
\def \Ds {D\kern-.7em / \kern.5em}

\def \II {I\hspace{-.1em}I\hspace{.05em}}

\def \IIB {\mbox{\II B\hspace{.2em}}}

\def \ys {{y\kern-.5em / \kern.3em}}

\documentstyle[12pt]{article}

\def\[{\left [}
\def\]{\right ]}
\def\({\left (}
\def\){\right )}

\begin{document}
\begin{titlepage}
\begin{center}

\hfill   JHU-TIPAC-98009 \\
\hfill   {\tt hep-th/9809055 } \\
\hfill   September 1998

\vskip .5in

{\Large \bf A Note on AdS/SYM Correspondence \\ on the Coulomb Branch}

\vskip .7in

Yi-Yen Wu

\vskip .3in
{\em     Department of Physics and Astronomy \\
             Johns Hopkins University \\
             Baltimore, Maryland 21218, USA} 

\end{center}

\vskip 1.0in

\begin{abstract}
We study Maldacena's conjecture and the AdS/SYM correspondence on the Coulomb branch.  
Several interesting aspects of this conjectured AdS/SYM correspondence on the Coulomb 
branch are pointed out and clarified.
\end{abstract}
\vskip .5in

%

\end{titlepage}

\newpage
\renewcommand{\thepage}{\arabic{page}}
\setcounter{page}{1}
\setcounter{footnote}{0}

\section{Introduction}
One of the remarkable AdS/CFT dualities (Anti-de Sitter space/Conformal Field Theory) 
conjectured by Maldacena is the equivalence between ${\cal N}\!\!=\!\! 4$ four-dimensional 
supersymmetric Yang-Mills theory (SYM) and Type \IIB string theory on AdS$_5$$\times$S$^5$ 
\cite{Mald}. Prescriptions for testing this correspondence have been given in 
\cite{Gubser,Witten1}, and this conjecture has been tested in many ways. (See \cite{review} for 
a recent review.) Recently, there is also some evidence for this conjecture at finite $N$ 
\cite{Aharony}. AdS/SYM correspondence is natural from the point of view of holography 
\cite{Witten1,Hooft}, and another interesting aspect of this correspondence is the study of Type 
\IIB branes in AdS$_5$$\times$S$^5$ and its SYM correspondence 
\cite{string,Witten2,Gross,Banks,CHW}.\footnote{Recently there is also a study of Type \IIB 
branes in AdS$_5$$\times$X$_5$ and its ${\cal N}\!\!=\!\! 1$ SYM correspondence \cite{X5}, 
where X$_5$ is a five-dimensional Einstein manifold \cite{Gubser2}.}

In \cite{Mald}, the AdS/SYM correspondence on the Coulomb branch has also been suggested. 
Recently, Douglas and Taylor \cite{Douglas} proposed that D3-branes in the AdS$_5$ bulk are  
equivalent to an ${\cal N}\!\!=\!\! 4$ four-dimensional SYM on the Coulomb branch where the 
adjoint Higgs scalars have non-vanishing vacuum expectation values ({\em 
vev}'s).\footnote{Some recent studies related to \cite{Douglas} can be found in \cite{p1,p2}.} In 
this letter, we will look more closely at this proposal. In Section 2, we study the AdS/SYM 
correspondence on the Coulomb branch following Maldacena's original argument in \cite{Mald}. 
The relevance of Witten's study of D3-branes in the AdS$_5$ bulk \cite{Witten2} is emphasized. 
An interesting duality between D3-brane configurations in the AdS$_5$ bulk suggested by this 
conjectured  AdS/SYM correspondence on the Coulomb branch is also pointed out. In Section 
3, we analyze an interesting aspect of this AdS/SYM correspondence on the Coulomb branch. 
Closely related to Witten's study of D3-branes as domain walls \cite{Witten2}, a consistency 
check at finite $N$ for the conjectured AdS/SYM correspondence on the Coulomb branch is 
also given by studying D3-branes in the AdS$_5$ bulk. (We note that Witten's argument for 
D3-branes in the AdS$_5$ bulk as domain walls of Type \IIB string theory on 
AdS$_5$$\times$S$^5$ in \cite{Witten2} also applies to M2-branes (M5-branes) in the AdS$_4$ 
(AdS$_7$) bulk, and they are therefore domain walls of M-theory on AdS$_4$$\times$S$^7$ 
(AdS$_7$$\times$S$^4$). As we will see, this suggests that many of the arguments in this letter 
can be extended to the study of M2-branes (M5-branes) in the AdS$_4$ (AdS$_7$) bulk for 
M-theory on AdS$_4$$\times$S$^7$ (AdS$_7$$\times$S$^4$).)

\section{AdS/SYM Correspondence on the Coulomb Branch}

\subsection{Argument $\grave{a}$ $la$ Maldacena and the Conjecture}
The D3-brane in AdS$_5$$\times$S$^5$ considered throughout this letter is {\it  a D3-brane 
unwrapped on S$^{\,5}$ and parallel to the AdS$_5$ boundary}. That is, it sits at a point of 
S$^5$ and its world volume fills the four non-radial directions of AdS$_5$. According to Witten 
\cite{Witten2}, $\Sigma_{(5)}$ changes by $\pm 1$ unit as we cross this D3-brane, where 
$\Sigma_{(5)}$ is the flux of R-R 5-form field strength $F_{(5)}$ over S$^5$. 
$\Sigma_{(5)}\rightarrow\Sigma_{(5)}\pm 1$ corresponds to two kinds of D3-branes, where the 
$F_{(5)}$'s supported on them are opposite in sign. A naive question is whether and how these 
D3-branes are stable\footnote{In this letter, ``stable'' means stable against the AdS gravitational 
force.} in AdS$_5$? In the study of mesons and baryons as strings and wrapped 5-branes in 
AdS$_5$ \cite{string,Witten2,Gross,Israel}, the AdS gravitational force acting on the string and 
wrapped 5-brane is balanced by the tension of the string(s) ending on the AdS$_5$ boundary, 
and therefore these configurations are stable. Unlike other branes, D3-branes carry R-R 
five-form field strength $F_{(5)}$. This $F_{(5)}$ interacts with the background $F_{(5)}$ of 
AdS$_5$$\times$S$^5$, which results in a R-R force acting on the D3-brane. These D3-branes 
are of two kinds: their $F_{(5)}$'s are opposite in sign. It is natural to expect that, for one kind of 
the D3-branes, the R-R force cancels the AdS gravitational force, and therefore it is stable. For 
the other kind of the D3-branes whose $F_{(5)}$ is opposite in sign to that of the former, the R-R 
force adds to the AdS gravitational force; it is unstable and accelerates toward the AdS$_5$ 
horizon. The former is exactly the analogy of cancellation of NS-NS and R-R forces between two 
parallel D3-branes in flat ten-dimensional spacetime, and the latter is analogous to the instability 
of parallel anti-D3-brane and D3-branes.\footnote{It is natural to refer to the above stable 
D3-brane as ``D3-brane in AdS$_5$'', and the above unstable D3-brane as ``anti-D3-brane in 
AdS$_5$''. However, here we simply call them stable and unstable D3-branes in AdS$_5$.} For  
the AdS/SYM correspondence on the Coulomb branch, naturally the main concern will be 
stable D3-branes in the AdS$_5$ bulk.

In the following, we try to study the AdS/SYM correspondence on the Coulomb branch by 
``deriving'' this correspondence following Maldacena's argument \cite{Mald}. To avoid 
unnecessary overlap with \cite{Mald}, the notations and conventions of \cite{Mald} are always 
assumed whenever it is possible. We begin with $N$ D3-branes in flat ten-dimensional 
spacetime. Consider the case where we have two groups of parallel D3-branes, $(N\!-\!M)$ 
D3-branes and $M$ D3-branes, separated by $\vec{r}$. If we start with Type \IIB string theory 
and take the decoupling limit $\alpha'\rightarrow 0$ while holding 
$\vec{W}$=$\,\vec{r}$/$\alpha'$ fixed, we obtain an ${\cal N}\!\!=\!\! 4$ four-dimensional $U(N)$ 
SYM with gauge group $\,U(N)\,$ spontaneously broken to $\,U(N\!-\!M)$$\times$$U(M)\,$ by 
Higgs scalar {\em vev}'s in the adjoint representation of $U(N)$, where the $N$ eigenvalues of 
Higgs scalar {\em vev}'s parametrize the positions of $N$ D3-branes.

Next, consider the supergravity (SUGRA) solution of the above D3-brane configuration and take 
the same decoupling limit. The resulting metric is
\begin{eqnarray} \label{metric}
\frac{ds^{2}}{\alpha'}&=&
\frac{U^2}{\sqrt{4\pi g\(N-M+\frac{MU^4}{|\vec{U}-\vec{W}|^4}\)}}dx^{2}_{\|} \;+\;
\frac{\sqrt{4\pi g\(N-M+\frac{MU^4}{|\vec{U}-\vec{W}|^4}\)}}{U^2}dU^2
\nonumber \\     & &  \hspace{4.9cm}   \;+\;
\sqrt{4\pi g\(N-M+\frac{MU^4}{|\vec{U}-\vec{W}|^4}\)}d\Omega_{5}^{2},
\end{eqnarray}
where $U$=$|\vec{U}|$, $\vec{U}/U$ is a point on S$^5$, and $d\Omega_{5}$ is the volume 
element of S$^5$. According to \cite{Mald}, the resulting theory in the SUGRA approximation is 
Type \IIB SUGRA on the background (\ref{metric}). Furthermore, we should ask what the 
underlying theory is beyond the SUGRA approximation. As we will argue, the background 
(\ref{metric}) represents the SUGRA solution describing $M$ stable D3-branes of Type \IIB string 
theory on AdS$_5$$\times$S$^5$. Therefore, the underlying theory should be Type \IIB string 
theory on AdS$_5$$\times$S$^5$ with $M$ stable D3-branes in the AdS$_5$ bulk.

Now we argue that the static background (\ref{metric}) is the SUGRA solution describing $M$ 
stable D3-branes of Type \IIB string theory on  AdS$_5$$\times$S$^5$. Firstly, (\ref{metric}) 
solves Type \IIB SUGRA equations of motion trivially. Secondly, as we move from 
$U\!\!=\!\!\infty$ to $U\!\!=\!\!0$, the geometry of (\ref{metric}) changes from 
$\,$(AdS$_5$$\times$S$^5$)$_{N}$$\,$ to $\,$(AdS$_5$$\times$S$^5$)$_{N\!-\!M}$. This 
corresponds to Type \IIB string theory on AdS$_5$$\times$S$^5$, where 
$\Sigma_{(5)}$=$(N\!-\!M)$ at $U\!\!=\!\!0$ and $\Sigma_{(5)}$=$N$ at $U\!\!=\!\!\infty$, and 
therefore $\Sigma_{(5)}$ changes by $M$ units across the AdS$_5$ bulk. According to Witten 
\cite{Witten2}, $\Sigma_{(5)}\rightarrow\Sigma_{(5)}+1$ when we cross a D3-brane along 
$U\!\!=\!\!\infty\!\rightarrow\!0$, and $\Sigma_{(5)}\rightarrow\Sigma_{(5)}-1$ when crossing a 
D3-brane whose $F_{(5)}$ is opposite in sign to that of the former. As argued in the beginning of 
this section, only one of them is stable; the other always accelerates toward $U\!\!=\!\!0$. We 
conclude that there must be $M$ stable D3-branes in the AdS$_5$ bulk. Thirdly, $\vec{W}$ is 
naturally identified as the position of $M$ D3-branes, where (\ref{metric}) is singular at 
$\vec{U}\!=\!\vec{W}$. Furthermore, note that the metric of (\ref{metric}) does not depend on 
$x_{\|}$, and therefore it does describe D3-branes unwrapped on S$^5$ and parallel to the 
AdS$_5$ boundary. Therefore we argue that the background (\ref{metric}) is indeed the Type 
\IIB SUGRA solution describing $M$ stable D3-branes of Type \IIB string theory on 
AdS$_5$$\times$S$^5$, where $\Sigma_{(5)}$=$(N\!-\!M)$ for $U<|\vec{W}|$ and 
$\Sigma_{(5)}$=$N$ for $U>|\vec{W}|$.\footnote{Consider in the decoupling limit \cite{Mald} the 
SUGRA solution for two groups of parallel M2-branes, $(N\!-\!M)$ M2-branes and $M$ 
M2-branes, separated by $\vec{r}$. Similarly we can argue that this SUGRA solution in the 
decoupling limit ($\vec{r}/l_{p}^{3/2}$ is fixed as $l_{p}\rightarrow 0$) describes $M$ stable 
M2-branes in the AdS$_4$ bulk for M-theory on AdS$_4$$\times$S$^7$. The observation that 
M2-branes in the AdS$_4$ bulk are domain walls of M-theory on AdS$_4$$\times$S$^7$ 
(see the last remark in \mbox{Section 1}; more precisely, (AdS$_4$$\times$S$^7$)$_{Z}$ 
$\rightarrow$ (AdS$_4$$\times$S$^7$)$_{Z\pm 1}$ 
across an M2-brane in the AdS$_4$ bulk) is again essential to this argument. The same 
consideration for M5-branes then leads to the SUGRA solution 
describing stable M5-branes in the AdS$_7$ bulk for M-theory on 
AdS$_7$$\times$S$^4$.} More precisely, according to Witten \cite{Witten2} this is Type \IIB 
string theory residing in the (AdS$_5$$\times$S$^5$)$_{N\!-\!M}$ string vacuum on one side of 
$M$ D3-branes, and residing in the (AdS$_5$$\times$S$^5$)$_{N}$ vacuum on the other side. 
The $M$ D3-branes are domain wall separating the two AdS$_5$$\times$S$^5$ vacua. We 
emphasize that, throughout this letter, this AdS string picture of Witten \cite{Witten2} is always 
assumed whenever we talk about D3-branes of Type \IIB string theory on 
AdS$_5$$\times$S$^5$ with definite $\Sigma_{(5)}$'s.

Following Maldacena's argument \cite{Mald}, the above argument leads to a conjectured 
AdS/SYM correspondence between field theory and string theory. The field theory is an ${\cal 
N}\!\!=\!\! 4$ SYM with gauge group $\,U(N)\,$ spontaneously broken to 
$\,U(N\!-\!M)$$\times$$U(M)\,$ by Higgs scalar {\em vev}'s in the adjoint representation of 
$U(N)$.\footnote{It has been argued that the SYM gauge group appropriate for AdS/SYM 
correspondence should be $SU(N)$ rather than $U(N)$, where the $U(1)$ part of $U(N)$ 
decouple \cite{Witten1,Aharony}. Throughout this letter we adopt a similar point of view that the 
$U(1)$ part of the boundary $U(N)$ SYM is ``frozen'' or ``non-dynamical'' \cite{probe2}. Only the 
remaining $SU(N)$ part of the $U(N)$ SYM is dynamical.} The string theory is Type \IIB string 
theory on AdS$_5$$\times$S$^5$, with $M$ stable coincident D3-branes in the AdS$_5$ bulk, 
where Type \IIB string theory resides in the (AdS$_5$$\times$S$^5$)$_{N}$ vacuum on the 
$U\!\!=\!\!\infty$ side of D3-branes and resides in the (AdS$_5$$\times$S$^5$)$_{N\!-\!M}$ 
vacuum on the $U\!\!=\!\!0$ side. Note that the presence of stable D3-branes in the AdS$_5$ 
bulk breaks the $SO(1,1)$ part of AdS$_5$ isometry group $SO(4,2)$ because of the 
$U$-position(s) of D3-branes. This corresponds to the fact that conformal symmetry of an ${\cal 
N}\!\!=\!\! 4$ SYM is broken because of the energy scale(s) introduced by non-vanishing Higgs 
scalar {\em vev}'s. This ``position/scale correspondence'' is a general feature of AdS/CFT(SYM) 
correspondence \cite{holo}, and has been observed in many examples 
\cite{CHW,Miao,probe1,probe2,Peet}.

It is straightforward to generalize the argument of this section to obtain the AdS/SYM 
correspondence for Type \IIB string theory on AdS$_5$$\times$S$^5$ with $M$ stable 
D3-branes in the AdS$_5$ bulk. These $M$ D3-branes consist of $K$ groups of coincident 
stable D3-branes. ($M=\sum_{i=1}^{K}M_{i}$. $M_{i}$ is the number of D3-branes in the $i$th 
group.) The relevant SUGRA solution is obtained by replacing 
\begin{equation}
\(N-M+\frac{MU^4}{|\vec{U}-\vec{W}|^4}\)\;\;\Rightarrow\;\;
\(N-M+\sum_{i=1}^{K}\frac{M_{i}U^4}{|\vec{U}-\vec{W_{i}}|^4}\)
\end{equation}
in (\ref{metric}), where $\,W_{1}\gg W_{2}\gg\cdots\gg W_{K}$. The corresponding field theory is 
an ${\cal N}\!\!=\!\! 4$ SYM with gauge group $\,U(N)\,$ spontaneously broken to 
$\,U(N\!-\!M)$$\times$$U(M_{1})$$\times\cdots\times$$U(M_{K})\,$ by Higgs scalar {\em vev}'s 
in the adjoint representation of $U(N)$.

\subsection{A Duality for Stable D3-Branes in AdS$_5$}
We discuss in this section an interesting implication of the conjectured AdS/SYM 
correspondence on the Coulomb branch. The argument $\,\grave{a}$ $la$ Maldacena in 
\mbox{Section 2.1} leads to the SUGRA solution (\ref{metric}) and an ${\cal N}\!\!=\!\! 4$ SYM 
with gauge group $\,U(N)\,$ spontaneously broken to $\,U(N\!-\!M)$$\times$$U(M)\,$ by Higgs 
scalar {\em vev}'s in the adjoint representation of $U(N)$. The SUGRA solution and the Higgs 
scalar {\em vev}'s of SYM are written explicitly as follows.\footnote{For convenience, here we 
use  $\vec{U}$ instead of $U$ and $\Omega_5$.}
\begin{eqnarray}
\frac{ds^{2}}{\alpha'}&=&
\frac{1}{\sqrt{4\pi g\(\frac{N-M}{|\vec{U}|^4} + \frac{M}{|\vec{U}-\vec{W}|^4}\)}}dx^{2}_{\|} \;+\;
\sqrt{4\pi g\(\frac{N-M}{|\vec{U}|^4}+\frac{M}{|\vec{U}-\vec{W}|^4}\)} d\vec{U}^2.  \\     
\left\langle\vec{{\bf X}}\right\rangle&=& \left( \begin{array}{cc}
                {\bf 0}_{N\!-\!M,N\!-\!M}  &   {\bf 0}_{N\!-\!M,M}    \\
                {\bf 0}_{M,N\!-\!M}  &  \vec{{\bf W}}_{\!\!M,M}
                    \end{array} \right).
\end{eqnarray}
$\vec{{\bf X}}$ denotes the six Higgs scalars in the adjoint of $U(N)$. $\vec{{\bf W}}_{\!\!M,M}$ is 
$\vec{W}$ times $M$$\times$$M$ identity matrix. As argued in \mbox{Section 2.1}, an ${\cal 
N}\!\!=\!\! 4$ $U(N)$ SYM with Higgs scalar {\em vev}'s specified by (4) corresponds to Type \IIB 
string theory on AdS$_5$$\times$S$^5$ ($\Sigma_{(5)}$=$N$ at $U\!\!=\!\!\infty$)\footnote{That 
is, near the AdS$_5$ boundary at $U\!\!=\!\!\infty$, Type \IIB string theory resides in the 
(AdS$_5$$\times$S$^5$)$_{N}$ vacuum. In the AdS$_5$ bulk, Type \IIB string theory may 
reside in different AdS$_5$$\times$S$^5$ vacua if there are stable D3-branes present 
\cite{Witten2}. Throughout this letter, this is what we mean by ``Type \IIB string theory on 
AdS$_5$$\times$S$^5$ ($\Sigma_{(5)}$=$N$ at $U\!\!=\!\!\infty$)''.} with $M$ stable D3-branes 
at $\vec{U}=\vec{W}$ in the AdS$_5$ bulk.

The same argument can be trivially repeated by another simple choice of coordinates, i.e., fixing
 the origin of coordinates $\vec{U}$ on the $M$ D3-branes instead of the $(N\!-\!M)$ D3-branes. 
This amounts to replacing $\vec{U}$ by $\vec{U}+\vec{W}$ in (3). The SUGRA solution and the 
Higgs scalar {\em vev}'s of SYM are:
\begin{eqnarray}
\frac{ds^{2}}{\alpha'}&=&
\frac{1}{\sqrt{4\pi g\(\frac{N-M}{|\vec{U}+\vec{W}|^4} + \frac{M}{|\vec{U}|^4}\)}}dx^{2}_{\|} \;+\;
\sqrt{4\pi g\(\frac{N-M}{|\vec{U}+\vec{W}|^4}+\frac{M}{|\vec{U}|^4}\)} d\vec{U}^2.  \\     
\left\langle\vec{{\bf X}}\right\rangle&=& \left( \begin{array}{cc}
                -\vec{{\bf W}}_{\!\!N\!-\!M,N\!-\!M}  &   {\bf 0}_{N\!-\!M,M}    \\
                {\bf 0}_{M,N\!-\!M}  & {\bf 0}_{M,M}
                    \end{array} \right).
\end{eqnarray}
Note that the SUGRA solutions (3) and (5) are related by a coordinate shift 
$\vec{U}$$\,\rightarrow\,$$\vec{U}+\vec{W}$. And the two $U(N)$ SYM's are related by a 
constant $U(1)$ shift, $\vec{{\bf X}}$$\,\rightarrow\,$$\vec{{\bf X}}+\vec{{\bf W}}_{\!\!N,N}$, in 
their Higgs scalar {\em vev}'s (4) and (6). $\vec{{\bf W}}_{\!\!N,N}$ is $\vec{W}$ times 
$N$$\times$$N$ identity matrix. As noted in \cite{probe2}, the $U(1)$ part of the boundary 
$U(N)$ SYM is ``frozen''. For Higgs scalars in the adjoint of $U(N)$, this ``frozen'' $U(1)$ is 
related to the fixing of coordinates in the SUGRA description. It is then clear that a coordinate 
shift $\vec{U}$$\,\rightarrow\,$$\vec{U}+\vec{W}$ in SUGRA corresponds to a constant $U(1)$ 
shift $\vec{{\bf X}}$$\,\rightarrow\,$$\vec{{\bf X}}+\vec{{\bf W}}_{\!\!N,N}$ in the Higgs scalars of 
boundary SYM. As argued in \mbox{Section 2.1}, an ${\cal N}\!\!=\!\! 4$ $U(N)$ SYM with Higgs  
scalar {\em vev}'s specified by (6) corresponds to Type \IIB string theory on 
AdS$_5$$\times$S$^5$ ($\Sigma_{(5)}$=$N$ at $U\!\!=\!\!\infty$) with $(N\!-\!M)$ stable 
D3-branes at $\vec{U}=-\vec{W}$ in the AdS$_5$ bulk.

The above arguments suggest an interesting duality between two D3-brane configurations: Type 
\IIB string theory on AdS$_5$$\times$S$^5$ ($\Sigma_{(5)}$=$N$ at $U\!\!=\!\!\infty$) with $M$ 
stable D3-branes at $\vec{U}=\vec{W}$, and Type \IIB string theory on AdS$_5$$\times$S$^5$ 
($\Sigma_{(5)}$=$N$ at $U\!\!=\!\!\infty$) with $(N\!-\!M)$ stable D3-branes at 
$\vec{U}=-\vec{W}$.\footnote{Therefore, it is a duality which formally takes $M$ to $(N\!-\!M)$, 
and $\vec{U}=\vec{W}$ to $\vec{U}=-\vec{W}$.} According to the above arguments or directly 
from the conjectured AdS/SYM correspondence on the Coulomb branch, the two $U(N)$ SYM's 
describing these two D3-brane configurations are simply related by a constant $U(1)$ shift, 
$\vec{{\bf X}}$$\,\rightarrow\,$$\vec{{\bf X}}\pm\vec{{\bf W}}_{\!\!N,N}$, in the Higgs scalars. 
Therefore, these two D3-brane configurations are ``dual'' to each other in the sense that they are 
related to each other simply by a coordinate shift 
$\vec{U}$$\,\rightarrow\,$$\vec{U}\pm\vec{W}$. 

This duality seems surprising from the point of view of AdS$_5$$\times$S$^5$ string theory. On 
the other hand, it can be regarded as a consequence of the conjectured AdS/SYM 
correspondence on the Coulomb branch.\footnote{It is also straightforward to extend this duality 
to D3-brane configurations which contain several groups of coincident stable D3-branes in the 
AdS$_5$ bulk.} The above arguments suggest how this duality arises. Whether or how this 
duality makes sense for AdS string theory remains to be seen.

\section{D3-Branes in AdS$_5$ and Coulomb Branch of SYM$_4$}
Together with the conjectured AdS/SYM correspondence on the Coulomb branch, 
the original AdS/SYM 
correspondence conjectured by Maldacena can be understood in a more general context: an 
${\cal N}\!\!=\!\! 4$ four-dimensional $U(N)$ SYM is dual to Type \IIB string theory on 
AdS$_5$$\times$S$^5$, where the R-R 5-form flux over S$^5$ is $\Sigma_{(5)}$=$N$ at 
$U\!\!=\!\!\infty$.\footnote{See Footnote 8.} Whether $U(N)$ is spontaneously broken by adjoint 
Higgs scalar {\em vev}'s corresponds to whether there are stable D3-branes in the AdS$_5$ 
bulk. There is an interesting aspect of this AdS/SYM correspondence from the point of view of 
SYM. Consider the ${\cal N}\!\!=\!\! 4$ $U(N)$ SYM for AdS/SYM correspondence. A Higgs 
scalar in the adjoint representation of $U(N)$ can have at most $N$ independent eigenvalues. In 
terms of the AdS/SYM correspondence in \mbox{Section 2.1}, this simple fact of SYM means 
that an ${\cal N}\!\!=\!\! 4$ $U(N)$ SYM corresponds to Type \IIB string theory on 
AdS$_5$$\times$S$^5$ ($\Sigma_{(5)}$=$N$ at $U\!\!=\!\!\infty$) with $M$ stable D3-branes in 
the AdS$_5$ bulk, and $M$ can only be $\;0,1,\cdots,(N\!-\! 1)$, depending on whether and how 
the $U(N)$ is broken by Higgs scalar {\em vev}'s. Therefore for Type \IIB string theory 
($\Sigma_{(5)}$=$N$ at $U\!\!=\!\!\infty$) with $M$ ($M$=$N$ or $M$$>$$N$) stable D3-branes, 
it seems that this AdS/SYM correspondence breaks down and there is no dual SYM description. 
The other possibility is that these string theory configurations with $M$$\ge$$N$ are ill defined. 
This seeming obstruction to $M$$\geq$$N$ is derived from SYM consideration alone. If the 
same obstruction can be understood by purely string/SUGRA consideration, it can be regarded 
as a consistency check at finite $N$ for the conjectured AdS/SYM correspondence on the 
Coulomb branch. Such a string/SUGRA analysis is possible by studying D3-branes in the 
AdS$_5$ bulk as follows.

There are two kinds of D3-branes in the AdS$_5$ bulk: $\Sigma_{(5)}\rightarrow\Sigma_{(5)}\pm 
1$ when we cross a D3-brane along $U\!\!=\!\!\infty\!\rightarrow\! 0$. As argued in \mbox{Section 
2.1}, only one of them is stable against AdS gravity. Which one is stable can be determined, for 
example, by a string theory calculation. However, for our purpose a simple SUGRA argument as 
follows will be sufficient. It has been argued that the static SUGRA solution (\ref{metric}) 
describes D3-branes of Type \IIB string theory on AdS$_5$$\times$S$^5$, where $\Sigma_{(5)}$ 
always decreases when we cross a D3-brane along $U\!\!=\!\!\infty\!\rightarrow\! 0$. Because 
only stable D3-branes have static SUGRA solution, the existence of (\ref{metric}) determines that 
$\Sigma_{(5)}\rightarrow\Sigma_{(5)}-1$ when we cross a stable D3-brane along 
$U\!\!=\!\!\infty\!\rightarrow\! 0$.\footnote{And $\Sigma_{(5)}\rightarrow\Sigma_{(5)}+1$ when we 
cross an unstable D3-brane along $U\!\!=\!\!\infty\!\rightarrow\! 0$.} This observation is essential 
to the following discussion.

Next, consider Type \IIB string theory on AdS$_5$$\times$S$^5$ ($\Sigma_{(5)}$=$N$ at 
$U\!\!=\!\!\infty$) with $M$ stable D3-branes in the AdS$_5$ bulk, where these $M$ D3-branes 
consist of $K$ separate groups of coincident stable D3-branes. ($M$=$\sum_{i=1}^{K}M_{i}$. 
$M_{i}$ is the number of D3-branes in the $i$th group ($i$=$1,2,\cdots,K$) indexed along 
$U\!\!=\!\!\infty\!\rightarrow\! 0$.) Firstly, consider $M$$<$$N$. According to Witten \cite{Witten2} 
and the above observation, the Type \IIB string vacuum jumps from one to another, 
$\;$(AdS$_5$$\times$S$^5$)$_{(N)}\;$ $\rightarrow$
$\;$(AdS$_5$$\times$S$^5$)$_{(N-M_{1})}\;$ $\rightarrow$ $\cdots$ $\rightarrow$
$\;$(AdS$_5$$\times$S$^5$)$_{(N-M_{1}\cdots -M_{K\!-\!1})}$$\;$ $\rightarrow$
$\;$(AdS$_5$$\times$S$^5$)$_{(N-M)}$, 
as we cross each group of coincident D3-branes along $U\!\!=\!\!\infty\!\rightarrow\! 0$. The $i$th  
group of D3-branes is a domain wall interpolating between the 
$\;$(AdS$_5$$\times$S$^5$)$_{(N-M_{1}\cdots -M_{i\!-\!1})}$$\;$ and 
$\;$(AdS$_5$$\times$S$^5$)$_{(N-M_{1}\cdots -M_{i})}$$\;$ string vacua. For $M$$<$$N$, the 
above spacetime is always of the anti-de Sitter type. According to the AdS holographic principle 
proposed in \cite{Witten1}, a boundary field theory description therefore should exist for 
$M$$<$$N$. This is consistent with the expectation of the conjectured AdS/SYM 
correspondence on the Coulomb branch.

Secondly, consider $M$=$N$. As we cross the last ($K$th) group of stable D3-branes along 
$U\!\!=\!\!\infty\!\rightarrow\! 0$, naively we expect that \mbox{$\Sigma_{(5)}$=$M_{K}$ 
$\,\rightarrow\,$ $\Sigma_{(5)}$=0}. Note that $\Sigma_{(5)}$=0 indicates that the spacetime is 
no longer of the anti-de Sitter type. Because the AdS holographic principle \cite{Witten1} has no 
natural generalization to non-AdS spacetime \cite{Witten3}, this suggests that the case of 
$M$=$N$ does not have a boundary field theory description in the sense of \cite{Witten1}. On 
the other hand, since the spacetime across the $K$th group of D3-branes is not of AdS type by a 
naive analysis, it may be not  even appropriate to talk about Type \IIB string theory on 
AdS$_5$$\times$S$^5$ ($\Sigma_{(5)}$=$N$ at $U\!\!=\!\!\infty$) with $M$=$N$ stable 
D3-branes in the AdS$_5$ bulk in the beginning. Thirdly, for Type \IIB string theory on 
AdS$_5$$\times$S$^5$ ($\Sigma_{(5)}$=$N$ at $U\!\!=\!\!\infty$) with $M$ ($M$$>$$N$) stable 
D3-branes in the AdS$_5$ bulk, the above consideration for $M$=$N$ also applies because 
$M$ ($M$$>$$N$) D3-branes always contain $N$ D3-branes as a subset.

In conclusion, the string/SUGRA analysis shows that only the cases of $M=0,1,2,\cdots,(N\!-\! 
1)$ stable D3-branes can have dual SYM descriptions. This obstruction to $M$$\ge$$N$ is 
exactly what is expected from purely SYM consideration. In this sense this string/SUGRA 
analysis can be regarded as a consistency check for the conjectured AdS/SYM 
correspondence on the Coulomb branch.

\section*{Acknowledgement}
I thank Chong-Sun Chu and Pei-Ming Ho for many useful discussions. This work  was supported 
in part by NSF grant No. PHY-9404057.

\end{document}